\documentclass[11pt]{article}
\usepackage{amssymb,palatino}

\begin{document}

\newcommand{\ea}{et al.}
\newcommand{\be}{\begin{equation}}
\newcommand{\ee}{\end{equation}}
\baselineskip 21pt
\parindent 10pt
\parskip 9pt

\noindent  SHORT COMMUNICATION to the {\em Journal of Biomechanics}\hfill 7/2009

 \vskip 0.2in
\noindent {\large\bf Scaling of human body mass with height: the Body Mass Index revisited}\\[0.1in]
\noindent {\large\em N. J. MacKay}

\noindent Department of Mathematics, University of York, York YO10
5DD, UK \\ email: nm15@york.ac.uk

\vskip 0.2in
\noindent\parbox[t]{5in}{\small {\bf ABSTRACT}: We adapt a biomechanical argument of Rashevsky, which places limits on the stress experienced by a torso supported by the legs, to deduce that body mass $m$ of growing children should scale as the $p$th power of height $h$ with $7/3<p<8/3$. Further arguments based on stability and heat loss suggest that $p$ should be close to $8/3$. The arguments are extended to suggest that waist circumference $w$ should scale as $h^q$ with $q$ near the lower end of $2/3\leq q \leq 1$. Data from Hong Kong and British children are consistent with these hypotheses.\\[0.1in] {\bf Keywords:} allometric scaling; Body Mass Index}

\vspace{0.2in} \noindent{\large\bf 1. Introduction}

Considered in the light of scaling arguments, the Body Mass Index (BMI; Keys \ea, 1972), an individual's mass $m$ (in kilograms) divided by the square of his or her height $h$ (in metres), is intriguing. First, it is not scale-invariant: if two individuals have the same body shape, but one is $r$ times taller and thus of $r^3$ times greater mass than the other, then the taller's BMI is $r$ times greater. If BMI is to be a predictor of health and longevity, the implication is that, for a given body shape, taller people are less healthy---but then this may well be so (Samaras \ea, 2002). Such considerations become stark if we compare humans in all stages of development: a two-metre man with the body shape of a toddler is probably very unhealthy indeed.

If mass scales as the cube of height ($m\sim h^3$), it is easy to think of physical characteristics which might be related to health and which scale like $mh^{-2}$. Waist circumference is then proportional to BMI; and waistline is perhaps a better indicator of health than BMI (Janssen \ea, 2004).  Blood pressure would scale as the height of the effective column over our feet and thus like BMI, and indeed they are correlated (Holland \ea, 1993).

But mass does not scale as the cube of height. Is there some other value  $p \neq 3$ for which $m\sim h^p$ ?  All modern work is effectively confirmation and refinement of the observations of Quetelet (in French 1832-35; in English 1842), that for growing children $p\simeq2.5$ while among adults $p\simeq 2$. For adults, the current state of empirical understanding, distinguishing fat, muscle and bone, may be found in Heymsfield \ea, 2007 (and references therein). The key conclusion is that (despite variation in muscularity) levels of fat are maximally correlated with $mh^{-p}$ when $p\simeq2$. For developing children, data sets typically show allometric scaling ({\em i.e.\ }$p$ deduced from a good regression line of the logarithmic data) for ages five and up, with $p$ in the range $2.5$--$2.8$ (see for example Shiner \ea, 2001).

However, an explanation of these facts is lacking. The purpose of this note is to suggest one by adapting an old biomechanical argument of Rashevsky (1938, 1955), originally applied to quadrupeds. The initial argument predicts $p$ in the range $7/3\leq p \leq 8/3$. We then suggest some reasons why $p$ might lie at the upper end of this range. Thus our model matches observations for children rather than adults---but this is quite reasonable, for scaling arguments have merit only where the range of the data is large. Finally we comment on the scaling of waist circumference, and the implications of the argument for adults. We test our ideas with simple regression analysis of the logarithmic data from British and Hong Kong school-children: the results are consistent with our suggestions.

\vspace{0.2in} \noindent{\large\bf 2. The biomechanical argument}

Rashevsky (1938, 1955, echoed by McMahon, 1973) modelled quadrupeds as beams supported front and back, and deduced the scaling of mass with body length by imposing limits on  body `sag', the stress-induced curvature. Let us instead imagine a human torso as a cuboid block of breadth $x$, depth $y$ and height $z$. It is supported on two hip joints a distance $x$ (or at least proportional to $x$) apart. We will derive a scaling of torso mass $m_T$ (assumed proportional to torso volume $xyz$) with torso height $z$.

\vspace{0.1in} \noindent 2.1. {\bf Height}

Standard results for elastic beams imply that the sag $s$ at the mid-point between the hips is
$$ s \propto {F x^3 \over I_A}\,,$$
where $F$ is the downward force, here that of gravity and so proportional to $xyz$, and $I_A$ is the area moment of inertia, proportional to $yz^3$. We then have
\be s \sim {x^4yz\over yz^3}={x^4\over z^2}\ee -- note that this is independent of $y$. Now suppose that there is a maximum physiologically-acceptable tendency to sag, imposing
\be\label{sag}{s\over x}\leq K\ee for some fixed $K$.
Then, at the limit, $x^3 \sim z^2$ or $x \sim z^{2/3}$. This is the key scaling: it implies that
\be m_T \sim z^{5/3}y.\ee However, the argument tells us nothing about how $y$ scales with $x$ and $z$. The extreme possibilities are: \\ \centerline{
\parbox[t]{3.2in}{  $\qquad y\sim x,\quad$ so that $\quad m_T\sim x^2z\sim z^{7/3},\quad$ or }} \centerline{
\parbox[t]{3.2in}{$\qquad y\sim z,\quad$ so that $\quad m_T\sim xz^2 \sim z^{8/3}$.}}\\[10pt]
Thus, if we assume that overall mass $m\sim m_T$ and that overall height $z\sim h$, we have $m\sim h^p$ for $p$ in the range $2{1\over 3}$--$2{2\over 3}$. We will parametrize this range by assuming that $y$ scales principally as $y\sim z (x/z)^\alpha$, with $0\leq \alpha \leq 1$. Then $p={8-\alpha\over 3}$.

\vspace{0.1in} \noindent 2.2. {\bf Torso depth}

Can we further deduce the scaling of $y\,$? ---that is, determine $\alpha$ and thereby $p\,$?
Suppose we believe that humans have evolved for maximum adult bipedal stability. This amounts to requiring that, for a given mass and thus fixed $xyz$, the mass moment of inertia $I_M$ about an axis through the hip joints be minimized as a function of $y$. Assuming a roughly uniform density for the human torso (for although the true density is higher in the abdomen, this does not affect the implication),\be
I_M\propto\int_{-y/2}^{y/2} \int_0^z  (y^2+z^2)\,dz\,dy = {yz\over 3}\left( {y^2\over 4}+z^2\right)\,.\ee   This is minimized at the (fixed) ratio $y/z=2$, which suggests that $y$ should increase as rapidly as possible, $y\sim z$ and thus $\alpha=0$, $p=8/3$, with $m\sim z^{8/3}$.

We might also consider heat loss. It is straightforward to show (we do not give details) that the ratio of torso surface area to mass decreases as mass increases, but does so most rapidly -- {\em i.e.\ }adults' capacity to conserve heat is maximized -- when $y$ increases most rapidly with $z$: that is, again, $y\sim z$, $\alpha=0$ and $p=8/3$. It is interesting to wonder whether the desirability of this differs between races, and whether this affects differences in BMI and its implications (Calle \ea, 1999; Ruff, 2002).

For a first test of our hypothesis we used data from schoolchildren in Hong Kong (aged 6-18; tables 1 \& 2 of So \ea, 2008) and Britain (aged 5-17; tables 1 \& 2 of McCarthy \ea, 2001), and performed simple regression analysis on the logarithmic data, binned by year of age. (For a full analysis one should certainly disaggregate this data, but the year-bins are sufficient for our preliminary purposes.) We performed standard checks on the binned data, and excluded from the British data the lowest year of age, which for each model had a Cook's distance greater than one ({\em i.e.\ }undue influence on the result). For boys, the Hong Kong data from 1963, 1993 and 2005-6 yielded respectively $p=2.69\pm0.07,\, 2.68\pm0.04$ and $2.67\pm0.05$ (with all multiple-$R^2\simeq0.99$). The British data gave $p=2.70\pm0.05$ (with $R^2>0.99$). This is all strikingly consistent with the ideas above.  For girls, the Hong Kong figures were $2.88\pm0.10,\, 2.93\pm0.07$ and $2.90\pm0.04$, while the British data gave $p=2.89\pm0.08$, all rather less so. In mitigation, we note that birthing constraints have been at least as strong as structural ones in the evolution of the female pelvis (Stewart, 1984), and lead to greater hip separation and thus larger $p$.

\vspace{0.1in} \noindent 2.3. {\bf Waist circumference}

We now consider waist circumference $w$, which we assume to scale principally as $x(y/x)^\beta$ for some unknown $\beta$ with $0\leq \beta \leq 1$ ---that is, somewhere between $w\sim x$ and $w\sim y$. Then $w\sim z^q$, where
\be w\sim x \left( {y\over x}\right)^\beta \sim z^{{2\over 3}(1-\beta)} \left\{ z \left({x \over z}\right)^\alpha \right\}^\beta
\ee and thus $q = 2/3 + \beta/3 -\alpha\beta/3$. Our na\"ive rectangular torso has a waistline of $w=2x+2y$, but its shortcomings as a model are  clear. An elliptical torso with breadth $x$ and depth $y$, with typical values of roughly $y/x\simeq 2/3$, would have $w\simeq 2x+y$ (computed by linear approximation to the appropriate elliptic integral). So we can expect the scaling of $w$ to be dominated by $x$ rather than $y$, with perhaps $\beta\simeq 1/4$. With $\alpha\simeq 0$ from our earlier arguments, we expect $2/3\leq q \leq 1$, and towards the lower end of the range. With $\beta\simeq 1/4$  we might expect $q\simeq 3/4$.

Again we can test this hypothesis. Table 2 of Sung \ea\ (2008) uses a variant of the Hong Kong 2005-6 data (yielding $p=2.66\pm0.04$ for boys and $2.93\pm0.06$ for girls). Fitting mean waist circumference to $h^q$ gives $q=0.68\pm0.04$ (boys; $R^2>0.96$) and $0.71\pm0.02$ (girls; $R^2>0.99$). The UK data gave $q=0.82\pm0.02$ (boys) and $0.70\pm0.03$ (girls), both with $R^2\simeq0.99$.

\vspace{0.1in} \noindent 2.4. {\bf Adults}

We have not explained why, although developing children typically have $p>2.5$, among adults rather $p\simeq 2$. As we noted earlier, however, scaling arguments have less merit where the range of the data is small, as here. In the derivation above, for $p=2$ we would need the cross-sectional torso area to scale as $z$, rather than the na\"ive $z^2$ or the $z^{7/3}$--$z^{8/3}$  found above. Why might this be so? Note that in our model $y$ typically increases more rapidly with increasing height than does $x$, suggesting that taller individuals have thicker torsos, and presumably a stronger bound (\ref{sag}) as $z$ increases. We do not have a biomechanical argument for this, but suppose that $K\sim z^{-4/5}$ and $y\sim x^{3/2}$. Then we would have $x\sim z^{2/5}$, $y\sim z^{3/5}$ and $p=2$.

Whether or not an argument along such lines is correct, the bound on $x$ and constraints on $y$ will certainly have to be relaxed to explain variation among adults. But then this might easily be so, with both superseded by other optimal-evolution arguments (Vaughan, 2003).

\vspace{0.2in} \noindent  {\large\bf 3. Concluding remarks}

For growing boys, at least, this model is strikingly consistent with the facts. A more detailed study would ideally look not just at the individual data at a single time, but rather at longitudinal data, of the growth patterns over a decade of many individuals.

For girls and women, we would hardly expect our argument to take precedence over childbearing constraints, and so are unsurprised to find larger $p$. But to explain $p\simeq 2$ among adult men some more subtle argument is surely needed.

\vskip 0.1in
\noindent{\bf Acknowledgments}\\
I should like to thank Samer Kharroubi for comments on the data analysis.

\parindent 0pt
\parskip 6pt
\baselineskip 16pt

\pagebreak
{\large\bf References}

Calle E E, Thun M J, Petrelli J M, Rodriguez C and Heath C W, 1999, {\em Body-Mass Index and mortality in a prospective cohort of US adults}, New England J. Medicine {\bf 341}(15), 1097-1105

%Farfan H F, 1978, {\em The biomechanical advantage of lordosis and hip extension For upright activity: Man as compared with other anthropoids},
%Spine {\bf 3}(4), 336-342%

Heymsfield S B, Gallagher D, Mayer L, Beetsch J and Pietrobelli A,  2007, {\em Scaling of human body composition to stature: new insights into Body Mass Index}, American J. Clinical Nutrition {\bf 86}, 82-91

Holland F J, Stark O, Ades A E and Peckham C S, 1993, {\em Birth weight and Body Mass Index in childhood, adolescence and adulthood as predictors of blood pressure at age 36}, J. Epidemiology and Community Health {\bf 47}, 432-435

Janssen I, Katzmarzyk P T and Ross R, 2004, {\em Waist circumference and not Body Mass Index explains obesity-related health risk}, American J. Clinical Nutrition {\bf 79}, 379-384

Keys A, Fidanza F, Karvonen M J, Kimura N and Taylor H L, 1972, {\em Indices of relative weight and obesity}, J. Chronic Diseases {\bf 25}, 329-343

McCarthy H D, Jarrett K V and Crawley H F, 2001, {\em The development of waist circumference percentiles in British children aged 5.0--16.9y}, European J. Clinical Nutrition {\bf 55}, 902-907

McMahon T, 1973, {\em Size and shape in biology}, Science {\bf 179}, 1201-1204

Quetelet M A, 1842, {\em A treatise on man and the development of his faculties}, Edinburgh, Chambers: Vol.2, Ch.2, sect.2; p66

Rashevsky N, 1938, {\em Mathematical Biophysics} Chicago, University of Chicago Press: Vol.2, Ch.24

Rashevsky N, 1955, {\em Organic form as determined by function}, Annals of the New York Academy of Science {\bf 63}, 442-453

Ruff C, 2002, {\em Variations in human body size and shape}, Annual Reviews in Anthropology {\bf 31}, 211-232

Samaras T T, Elrick H and Storms L H, 2003, {\em Is height related to longevity?}, Life Sciences {\bf 72}, 1781-1802

Shiner J S and Uehlinger D E, 2001, {\em Body mass index: a measure for longevity}, Medical Hypotheses {\bf 57}(6), 780-783

So H K, Nelson E A S, Li A M, Wong E M C, Lau J T F, Gulden G S, Mak K H, Wang Y, Fok T F and Sung R Y T, 2008, {\em Secular changes in height, weight and Body Mass Index in Hong Kong children}, BMC Public Health {\bf 8}, 320

Stewart DB, {\em The pelvis as a passageway. I. Evolution and adaptations},  British J. Obstetrics and Gynaecology {\bf 91}(7), 611-617

Sung R Y T, So H K, Choi K C, Nelson E A S, Li A M, Yiu J A T, Kwok C W L, Ng P C and Fok T F, 2008, {\em Waist circumference and waist-to-height ratio of Hong Kong Chinese children}, BMC Public Health {\bf 8}, 324

Vaughan C L, 2003, {\em Theories of bipedal walking: an odyssey}, J. Biomechanics {\bf 36}(4), 513-523
\end{document}